%% file: conference_101719.tex
\documentclass[conference]{IEEEtran}
\IEEEoverridecommandlockouts
\usepackage[table,xcdraw]{xcolor}
\usepackage{cite}
\usepackage{amsmath,amssymb,amsfonts}
\usepackage{algorithmic}
\usepackage{graphicx}
\usepackage{textcomp}
\usepackage{xcolor}
\usepackage{balance}
\usepackage{tablefootnote}
\usepackage{diagbox}
\usepackage{eso-pic}
\usepackage[pdftex]{hyperref}
\usepackage{tikz}

\newcommand\copyrighttext{%
    \footnotesize \textcopyright 2022 IEEE. Personal use of this material is permitted.
    Permission from IEEE must be obtained for all other uses, in any current or future
    media, including reprinting/republishing this material for advertising or promotional
    purposes, creating new collective works, for resale or redistribution to servers or
    lists, or reuse of any copyrighted component of this work in other works.
    DOI: \href{https://doi.org/10.1109/BigData55660.2022.10020864}{https://doi.org/10.1109/BigData55660.2022.10020864}}
\newcommand\copyrightnotice{%
    \begin{tikzpicture}[remember picture,overlay]
        \node[anchor=south,yshift=10pt] at (current page.south) {\fbox{\parbox{\dimexpr\textwidth-\fboxsep-\fboxrule\relax}{\copyrighttext}}};
    \end{tikzpicture}%
}

\def\BibTeX{{\rm B\kern-.05em{\sc i\kern-.025em b}\kern-.08em
    T\kern-.1667em\lower.7ex\hbox{E}\kern-.125emX}}
\begin{document}

\newcommand{\dockeywords}{Scientific Workflow, Monitoring, Scientific Workflow Management System}

\title{Towards Advanced Monitoring \\ for Scientific Workflows}

\author{
    %
    %
        \IEEEauthorblockN{
        Jonathan Bader\IEEEauthorrefmark{1}, 
        Joel Witzke\IEEEauthorrefmark{2},
        Soeren Becker\IEEEauthorrefmark{1},
        Ansgar Lößer\IEEEauthorrefmark{3}, \\
        Fabian Lehmann\IEEEauthorrefmark{4}, 
        Leon Doehler\IEEEauthorrefmark{1},
        Anh Duc Vu\IEEEauthorrefmark{4},
        Odej Kao\IEEEauthorrefmark{1}}

        \IEEEauthorblockA{
            \IEEEauthorrefmark{1}
            \{firstname.lastname\}@tu-berlin.de,
            Technische Universität Berlin, Germany
        }
        \IEEEauthorblockA{
            \IEEEauthorrefmark{2}
            witzke@zib.de,
            Zuse Institute Berlin, Germany
        }
        \IEEEauthorblockA{
            \IEEEauthorrefmark{3}
            ansgar.loesser@kom.tu-darmstadt.de,
            TU Darmstadt, Germany
        }
        \IEEEauthorblockA{
            \IEEEauthorrefmark{4}
            \{fabian.lehmann,vuducanh\}@informatik.hu-berlin.de,
            Humboldt-Universität zu Berlin, Germany
        }
  
    }

\maketitle
\copyrightnotice

\begin{abstract}
\input{sections/0_abstract}
\end{abstract}


\section{Introduction}\label{sec:INTRO}
\input{sections/1_introduction}

\section{Background \& Related Work}\label{sec:Background_RW}
\input{sections/2_background_rw}

\section{Monitoring Layers}\label{sec:taxonomy}
\input{sections/3_taxonomy}

\section{Monitoring in existing SWMS}\label{sec:classification_swms}
\input{sections/4_swms}

\section{Conclusion}\label{sec:conc}
\input{sections/5_conclusion}

\section*{Acknowledgment}
{\small Funded by the Deutsche Forschungsgemeinschaft (DFG, German Research Foundation) as FONDA (Project 414984028, SFB 1404).}

\bibliographystyle{IEEEtran}
\balance
\bibliography{references}

\end{document}

%% file: sections/0_abstract.tex
Scientific workflows consist of thousands of highly parallelized tasks executed in a distributed environment involving many components.
Automatic tracing and investigation of the components' and tasks' performance metrics, traces, and behavior are necessary to support the end user with a level of abstraction since the large amount of data cannot be analyzed manually.
The execution and monitoring of scientific workflows involves many components, the cluster infrastructure, its resource manager, the workflow, and the workflow tasks.
All components in such an execution environment access different monitoring metrics and provide metrics on different abstraction levels. 
The combination and analysis of observed metrics from different components and their interdependencies are still widely unregarded.

We specify four different monitoring layers that can serve as an architectural blueprint for the monitoring responsibilities and the interactions of components in the scientific workflow execution context.
We describe the different monitoring metrics subject to the four layers and how the layers interact.
Finally, we examine five state-of-the-art scientific workflow management systems (SWMS) in order to assess which steps are needed to enable our four-layer-based approach.

%% file: sections/1_introduction.tex
In natural sciences, such as bioinformatics, material science, or earth observation, scientific workflows are used to automate the analysis of experiments and the collection of results~\cite{garcia2020sarek,yates2021reproducible,baderStypRekowski2022ICDMW, rettelbach2021quantitative,rettelbachGraphsSSDBM,baderReshi2022IPCCC, lehmann2021force, schaarschmidt2021workflow}.
Scientists are faced with systems becoming increasingly complex and data easily exceeding hundreds of gigabytes or even terabytes~\cite{heidsieck2019adaptive, vivian2017toil}.
To tackle these challenges, scientific workflow management systems (SWMS) have emerged and help to compose data analysis tasks into workflows, orchestrate the execution, and manage various artifacts.

SWMS offer a simplified interface to define input and output data and consequently allow domain scientist to reuse existing scripts and programs without the need to rewrite code and to use (big data) APIs with a possibly steep learning curve. 
Additionally, in recent years, the adoption of cloud computing environments to deploy and execute workflows gained in prevalence over traditional (HPC) clusters, i.e., due to the offered scalability, pay-as-you-go model, or wide variety of available heterogeneous hardware.

In cloud computing, monitoring capabilities are essential for efficient, reliable, and performant data processing by combining and providing metrics, logs, and traces of different workload executions. 
Although multi-modal monitoring is applied for a wide variety of use cases in the big data domain,
monitoring SWMS frequently refers to an abstract view on the current or historical workflow status or consolidated resource utilization.
Even though some approaches intend to offer further monitoring functionalities for workflow systems with dedicated software or infrastructure systems, the combination and analysis of observed real-time metrics from different layers of the utilized infrastructure is -- to the best of our knowledge -- still not widely regarded in SWMS.

We believe, that extending the monitoring capabilities of current SWMS, enabling real-time observability and combining performance metrics obtained from the workflow management system with lower level infrastructure metrics is essential to further improve the reliability of workflow executions and accelerate the identification of problems in the environment.
For instance, monitoring performance metrics such as memory usage, I/O operations, and network utilization help to identify bottlenecks across the computational infrastructure that slow down the overall runtime, may require hardware upgrades, or might point to configuration problems and workflow inefficiencies.
Moreover, through monitoring resource usage data, task resource requirements can be estimated to allocate resources efficiently in an automated manner~\cite{witt2019learning, witt2019feedback,bader2021tarema,scheinert2022PeronaRI,bader2022RL} and simultaneously enable automated system tuning.

To pave the way for further research in this direction, in this paper, we provide an architectural blueprint for monitoring scientific workflows.
Our architectural blueprint consists of four layers that consider the corresponding monitoring subject and which metrics need to be retrieved from underlying layers to be able to provide the respective reports.
We distinguish between four different layers, the resource manager layer, the workflow layer, the machine layer, and the task layer.
Thereby, the monitored metrics differ in their granularity and monitoring frequency.
We describe each layer's monitoring features and their dependencies on the underlying layers.

Finally, we investigate how a set of widely adopted SWMS employ monitoring capabilities, classify these into our four-layer-based monitoring blueprint, and carve out open challenges.

%% file: sections/2_background_rw.tex
In this section, we describe scientific workflows and examine workflow monitoring in detail.
\subsection{Scientific Workflows}

\begin{figure}[tb!]
\centering
    \includegraphics[width=0.8\columnwidth]{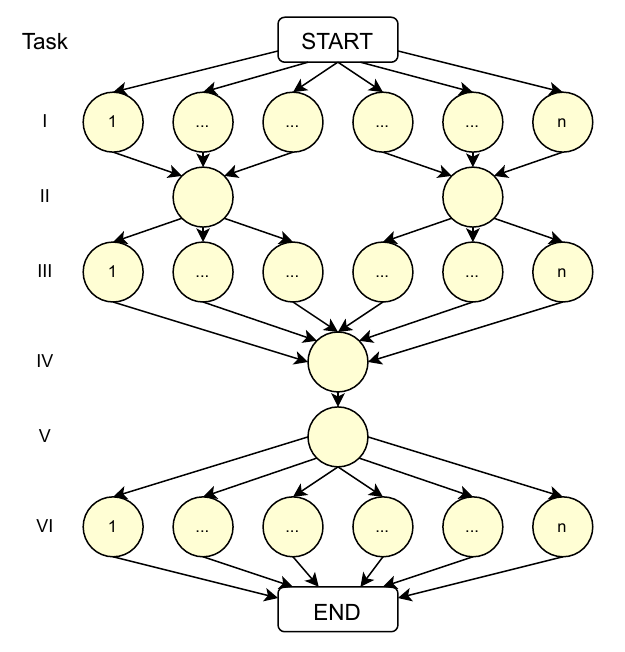}
    \caption{A directed acyclic graph representing a workflow with six different task definitions where task I, II, III, and VI spawn multiple instances in parallel, while task IV and V result in a single task instance.}
    \label{fig:workflow_example}
\end{figure}

Scientific workflows consist of black-box tasks~\cite{witt2019feedback,witt2019predictive,bader2021tarema, baderLotaruLocallyEstimating2022} that transform input data to output data for subsequent tasks.
Workflows are typically organized as directed acyclic graphs (DAG), see Figure~\ref{fig:workflow_example}, where nodes represent tasks and edges represent dependencies between the tasks and thus define the execution order.
The output from preceding tasks serves as input for a subsequent task.
Given such a dependency, a predecessor task has to finish successfully before the successor task can start processing. 
Figure~\ref{fig:workflow_example} shows a DAG representing a workflow where multiple instances of a task are executed in parallel for Task~I.
Task~II partly merges the results, whereby Task~III again performs a parallel execution of multiple instances.
Task~IV merges all results in a single task instance followed by another non-parallelizable task.
The last task is again executed in parallel task instances.

Typically, a workflow's level of parallelism depends on the number of input files, e.g., genomes, images, or materials to be analyzed.
Due to the black-box property of tasks, the system responsible for running the scientific workflows, the scientific workflow management system (SWMS), has no insight into the tasks. 
Therefore, tasks can be written in an arbitrary programming language and solely have to provide inputs and outputs according to defined standards.
However, scientists can easily reuse their existing scripts and programs as tasks instead of completely rewriting their existing code to build a workflow.

Scientific workflows are frequently executed on cluster infrastructures supervised by resource managers that allocate tasks to cluster nodes.
The monitoring component of such a resource manager supervises coarse hardware utilization, task executions, and helps to enforce the resource limits.

\subsection{Workflow Monitoring}
The monitoring of scientific workflows is considered from different perspectives in modern research.
A major focus in related work is the implementation of monitoring systems.  
Attempts are being made to implement an overarching solution independent of a respective software. 
For example, cloud-based services are being developed that allow the user to monitor the workflow in real time~\cite{rathnayake2018realtime}.

Mandal~et~al.~\cite{mandal2016toward} propose a prototype system for modeling and diagnosing run-time performance metrics of scientific workflows.
Thus, anomalies and sources of errors can be found during the execution of scientific workflows. 
The authors distinguish between workflow, application, and infrastructure monitoring. 
While workflow and application monitoring refer to the current status of the workflow, e.g., errors or state of execution, infrastructure monitoring examines the actual hardware metrics and parameters.
Although the authors propose a simple categorization of monitoring metrics, they focus on a system approach for end-to-end performance modeling and diagnosing instead of a general architectural blueprint for workflow monitoring.

In other research a more practical approach by attempting a more extensive classification is taken: 
Such an approach is proposed by Juve~et~al.~\cite{juve2015practical}, who want to achieve effective monitoring in high-throughput computing. 
Here, monitoring is classified into three monitoring mechanisms. 
These groups differ in accuracy, complexity, and overhead. 
With so-called interposed monitoring, an interface is created between a process and the consumed resource to observe in detail how the process is influencing it. 
Query-based monitoring uses data that the operating system stores and collects. 
However, the data quickly becomes outdated, leading to inaccuracies.
Lastly, so-called notification monitoring is used, which also uses data from the operating system.
However, here notifications are actively sent as soon as the status of a resource changes. 
In contrast to Juve~et~al.~\cite{juve2015practical}, we do not classify the monitoring mechanisms.
Instead, our architectural blueprint consists of layers that consider the corresponding monitoring subject and which metrics need to be retrieved from underlying layers in order to be able to provide the respective reports.

Tschueter~et~al.~\cite{tschuter2019top} provide another approach for a general monitoring classification.
They design a scheme that provides metrics from different workflow perspectives to support pinpointing bottlenecks.
This methodology consists of a workflow level, a job level, and a job step level.
On the workflow level, general performance data of the workflow -- including all jobs -- are considered. 
Here, the user can see the dependencies between jobs, the impact of a certain job, or which jobs form a bottleneck.
The job level only considers individual jobs and job steps, i.e., parts of a job.
Therefore, users can characterize selected jobs on this level and estimate how certain job steps influence the overall job performance.
Finally, the job step level provides the most detailed information in the form of traces.
The gathered performance data from detailed event logs support identifying the reason for a possible bottleneck.
Thus, parameters such as I/O are used to find the resource-intensive elements.
In contrast to the work by Tschueter~et~al., we take the execution infrastructure and its management into account.
Additionally, we assume a task (in the words of Tschueter~et~al. a job) to be atomic.
Therefore, we do not incorporate a job step.

Although the presented papers are the first steps toward a general architectural blueprint for efficient workflow monitoring, most approaches do not include the execution environment or allow for a reliable classification of monitoring data concerning scientific workflows.

%% file: sections/3_taxonomy.tex
\begin{figure}[]
\centering
    \includegraphics[width=0.8\columnwidth]{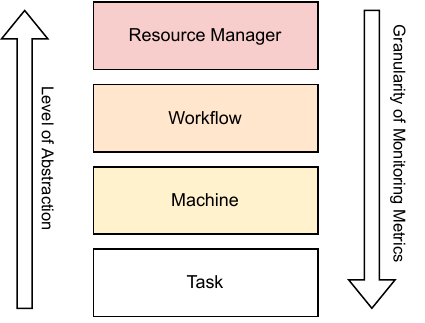}
    \caption{Hierarchical monitoring layers considering the abstraction level of a monitoring metric and the granularity of collected metrics.}
    \label{fig:pyramid}
\end{figure}

\input{tables/layers}

In this section, we first establish four monitoring layers, each defining the monitoring subject of the corresponding monitoring metric in the context of a scientific workflow execution.
Second, we explain the layers in detail and give examples of the monitored metrics for each of these layers.

\subsection{Overview}

Our architectural blueprint consists of four layers, the resource manager, the workflow, the machine, and the task layer, that consider the corresponding monitoring subject and which metrics need to be retrieved from underlying layers in order to be able to provide the respective reports.
The layers are selected according to the components in a scientific workflow execution environment as there are four logical distinctions.
Here, we explicitly do not follow a user-centric approach, i.e., an approach where the user interacts with the top layer, which is built on lower layers.
Figure~\ref{fig:pyramid} shows the layers and explains the relationship between the abstraction level of monitoring metrics and their granularity.
Higher layers are, in theory, able to retrieve the metrics from all underlying lower layers.
However, to operate, they only rely on parts of the underlying layers and, therefore, serve as a higher abstraction level.
For example, monitoring on the level of a resource manager, e.g., Slurm~\cite{slurm}, Kubernetes~\cite{kubernetes}, or HTCondor~\cite{condor}, is the highest abstraction level.
A resource manager does not need to know task-specific traces or low-level monitoring data from cluster machines such as syscalls but rather abstract metrics like task resource consumption and available resources on a machine.
The underlying workflow layer comprises metrics related to the abstract workflow definition, while the machine layer provides fine granular reports for the compute machine.
Lastly, the task layer monitoring provides fine granular task metrics and serves as the lowest layer in our hierarchy.
Table~\ref{tab:layers} provides an overview of the different layers and lists examples of monitored features from the underlying levels.
While some of the features are mandatory to ensure the workflow execution, e.g., the workflow representation feature, other features provide additional information.
Therefore, the presented table does not claim to be complete.

\subsection{Resource Manager Layer}

The resource manager supervises cluster machines and is responsible for assigning and executing submitted task instances on a machine.
Due to these responsibilities, the resource manager serves as the first layer and provides coarse monitoring information.
Therefore, the layer contributes summarized reports and statistics, e.g., the resource manager knows about all machines in the compute cluster, possibly concurrent running workflow executions, running and queued tasks, and the state information about the cluster's distributed file system.
The resource manager layer requires additional information from the workflow, machine, and task layers to perform assignments between tasks and machines.
Ideally, for workflow features, the resource manager retrieves several workflow information to handle the execution.
Examples are a workflow identifier to represent a certain workflow and to enable the execution of multiple workflows in parallel, a DAG status, and a DAG representation to ensure task dependencies.
The retrieved monitoring information from the cluster machines includes general information like a node's status, e.g., healthy, maintenance, or unhealthy, available and used amount of memory, disk, and CPU cores. 
Retrieved task information includes the task's status, requested and consumed resources, or task duration.

\subsection{Workflow Layer}

The second layer contains workflow metrics.
We assume that the workflow is submitted as a graph (frequently a DAG).
Therefore, the workflow describes the dependencies and restricts the order of task executions.
Each graph holds an identifier and can graphically be represented to the user, which increases the intelligibility of the underlying dependencies. 
Further, the workflow level delivers a status of the execution.
This is not restricted to an overall workflow flag like running or finished but also reports on the number of to-be-executed tasks in relation to finished ones, the progress of the overall execution, or failures in the workflow execution.
Additionally, features like the workflow makespan, the runtimes of the finished tasks, the submission time, or the number of previous workflow executions are reported.

\subsection{Machine Layer}

The machine layer includes all monitoring metrics related to a single compute machine in the cluster.
This includes coarse features that are frequently retrieved from higher levels like the number of available and used CPU cores or the amount of allocated and available memory.
However, at the specific machine layer, also more fine granular metrics are reported.
The machine type, e.g., bare-metal or virtualization is part of this reporting.
These metrics are extended with static and more fine-granular hardware characteristics, e.g., CPU architecture, CPU model, memory clock rates, disk partitions, or virtualization metrics.
Again, Table~\ref{tab:layers} lists some important monitoring features for this layer.

\subsection{Task Layer}

The task layer is the lowest level.
It describes monitoring features related to a single task and has no access to the upper levels, e.g., the overall workflow or the resource manager.
Examples of monitored features are logs from the task, e.g., logs from different log levels like warning or info messages, time-series resource consumption for certain areas of the task, e.g., memory consumption of a method in the code, or more general statistics like the task's execution time or the status.
Beyond that, more fine-granular kernel specific metrics can be monitored for each task.
Examples are system calls a task does, tracing the block device's I/O, page cache accesses, or the time a CPU task spends waiting until CPU time is assigned.
Further, through aggregated task metrics, a user is able to conduct a failure diagnosis or the debugging of a task, e.g., understanding performance bottlenecks.

\subsection{Considerations}
\label{subs:considerations}

\begin{figure}[tb!]
\centering
    \includegraphics[width=\columnwidth]{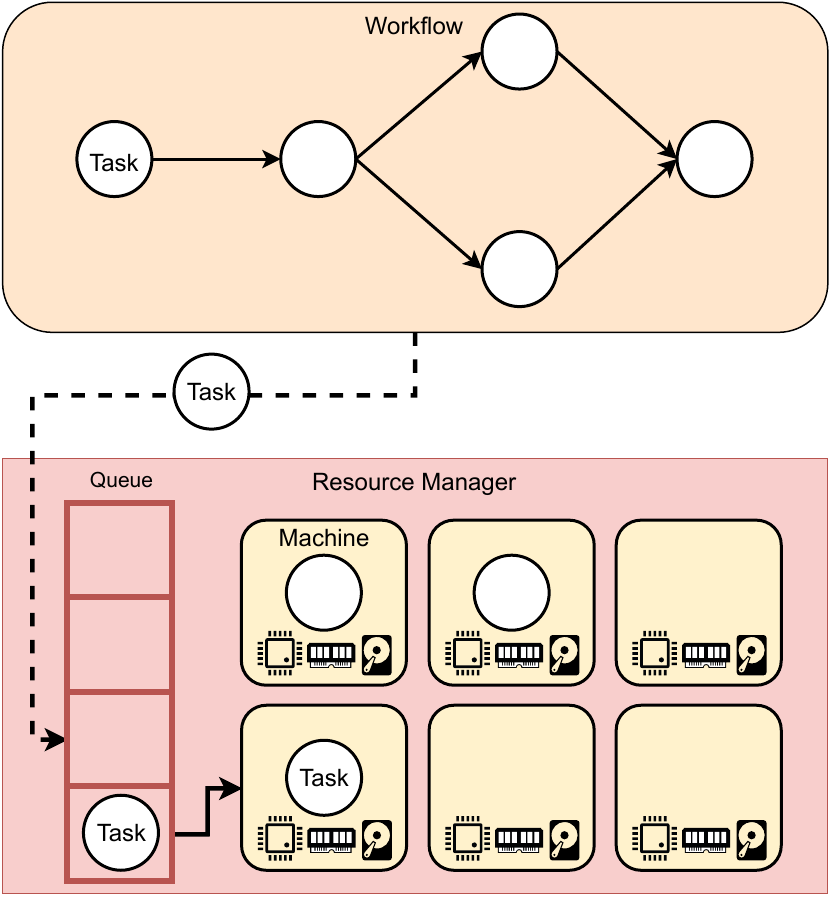}
    \caption{The figure describes how many state-of-the-art workflow management and resource managers interact from a monitoring perspective. Here, the workflow and the resource manager layer are disjoint, disabling the resource manager layer to retrieve workflow-specific monitoring metrics that would be mandatory for intelligent scheduling decisions.}
    \label{fig:dimensions_actual}
\end{figure}

The four layers present an ideal case where the resource manager is aware of the workflow and, therefore, the dependencies between the tasks.
Such an architecture allows for smart scheduling decisions since the resource manager can consider task dependencies.
However, many resource managers do not incorporate workflow structures and assume independently running tasks~\cite{rodrigo2017enabling, rodrigo2017scsf, bader2021tarema}.
In such a scenario, the workflow system submits ready-to-run instances to the resource manager's queue, and the resource manager runs a simple scheduling heuristic to assign a suitable machine~\cite{bader2021tarema, baderLotaruLocallyEstimating2022}.
Figure~\ref{fig:dimensions_actual} shows the relationship between the layers when the workflow layer is separated from the resource manager layer.
Here, the resource management layer is not able to access workflow specific metrics, e.g., DAG, or status, anymore.
Therefore, the hierarchy from Figure~\ref{fig:pyramid} does not hold.

Recently, there has been work towards enabling workflow-aware scheduling of HPC resource managers~\cite{rodrigo2017enabling, rodrigo2017scsf} that can be expressed by our presented monitoring hierarchy.

%% file: tables/layers.tex

\begin{table*}[]
\centering
\caption{The table shows the different monitoring layers, their corresponding monitoring features, and accessed features from lower monitoring layers.}
\begin{tabular}{|lcccc|}
\hline
\multicolumn{1}{|l|}{\diagbox{Monitoring Features}{Monitoring Layers} }                                                       & \multicolumn{1}{l|}{ \textbf{Resource Manager}} & \multicolumn{1}{l|}{\textbf{Workflow}} & \multicolumn{1}{l|}{\textbf{Machine}} & \multicolumn{1}{l|}{\textbf{Task}} \\ \hline
\multicolumn{5}{|l|}{\textbf{Resource Manager}}                                                                                                                                                                                                                                       \\ \hline
Infrastructure status                                                         & x                                                         &                                                   &                                                  &                                               \\
File system status                                                            & x                                                         &                                                   &                                                  &                                               \\
Running workflows                                                             & x                                                         &                                                   &                                                  &                                               \\ \hline
\multicolumn{5}{|l|}{\textbf{Workflow}}                                                                                                                                                                                                                                               \\ \hline
Status                                                                        & x                                                         & x                                                 &                                                  &                                               \\
Workflow specification                                                        & x                                                         & x                                                 &                                                  &                                               \\
Graphical representation                                                      &                                                           & x                                                 &                                                  &                                               \\
Workflow ID                                                                   & x                                                         & x                                                 &                                                  &                                               \\
Execution report                                                              &                                                           & x                                                 &                                                  &                                               \\
Previous executions                                                           &                                                           & x                                                 &                                                  &                                               \\ \hline
\multicolumn{5}{|l|}{\textbf{Machine}}                                                                                                                                                                                                                                                \\ \hline
Status                                                                        & x                                                         &                                                   & x                                                &                                               \\
Machine type                                                                  & x                                                         &                                                   & x                                                &                                               \\
Hardware specification                                                        &                                                           &                                                   & x                                                &                                               \\
Available resources                                                           & x                                                         &                                                   & x                                                &                                               \\
Used resources                                                                & x                                                         &                                                   & x                                                &                                               \\ \hline
\multicolumn{5}{|l|}{\textbf{Task}}                                                                                                                                                                                                                                                   \\ \hline
Task status                                                                        & x                                                         & x                                                 & x                                                & x                                             \\
Requested resources                                                           & x                                                         & x                                                 & x                                                & x                                             \\
Consumed resources                                                            & x                                                         & x                                                 & x                                                & x                                             \\
\begin{tabular}[c]{@{}l@{}}Resource consumption\\ for code parts\end{tabular} &                                                           &                                                   &                                                  & x                                             \\
Task ID                                                                       & x                                                         & x                                                 & x                                                & x                                             \\
Application logs                                                              &                                                           &                                                   &                                                  & x                                             \\
Task duration                                                                 & x                                                         & x                                                 & x                                                & x                                             \\
Low-level task metrics                                                             &                                                           &                                                   &                                                 & x                                              \\ 
Fault diagnosis                                                               &                                                           &                                                   &                                                  & x                                             \\ \hline

\end{tabular}
\label{tab:layers}
\end{table*}

%% file: sections/4_swms.tex
In this section, we briefly present existing scientific workflow management systems (SWMS) that we investigated regarding their monitoring capabilities according to our proposed architectural blueprint.
We show the results of that investigation and discuss them regarding adaptions necessary to fulfill the four-layer-based approach.

\subsection{Workflow Systems}
\label{subs:wssys}

\begin{table*}[]
  \centering
  \caption{Monitoring capabilities of different scientific workflow management systems (SWMS).}
  \begin{tabular}{|l|ccccc|}
  \hline
  \multicolumn{1}{|l|}{\diagbox{Monitoring Features}{SWMS}} & \multicolumn{1}{l|}{\textbf{Pegasus}} & \multicolumn{1}{l|}{\textbf{Nextflow}} & \multicolumn{1}{l|}{\textbf{Airflow}} & \multicolumn{1}{l|}{\textbf{Snakemake}} & \multicolumn{1}{l|}{\textbf{Argo}} \\
  \hline
  \multicolumn{6}{|l|}{\textbf{Resource Manager}} \\
  \hline
  Infrastructure status &   &   & x &   & \\
  File system status    &   &   &   &   &   \\
  Running workflows     &   &   &   &   &   \\
  \hline
  \multicolumn{6}{|l|}{\textbf{Workflow}}       \\
  \hline
  Status    & x & x & x & x & x \\
  Workflow specification    & x & x & x & x & x \\
  Graphical representation  &   & x & x & x & x \\
  Workflow ID               & x & x & x & x & x \\
  Execution report          & x & x & x & x & x \\
  Previous executions       & x & x & x &   & x \\
  \hline
  \multicolumn{6}{|l|}{\textbf{Machine}}       \\
  \hline
  Status                   &   &   &   &   &   \\
  Machine type             &   &   &   &   &   \\
  Hardware specification   &   &   &   &   &   \\
  Available resources      &   &   &   &   &   \\
  Used resources           &   &   &   &   &   \\
  
  \hline
  \multicolumn{6}{|l|}{\textbf{Task}}           \\
  \hline
  Task status        & x & x & x & x &   \\
  Requested resources   &   & x &   & x & x \\
  Consumed resources    & x & x &   &   & x \\
  \shortstack{Resource consumption \\ for code parts} &   &   &   &   &   \\
  Task ID               & x & x & x & x & x \\
  Application logs      & x & x & x & x & x \\
  Task duration         & x & x & x & x & x \\
  Low-level task metrics        &   &   &   &   &   \\
  Fault diagnosis       & x &   &   & x &   \\
  
  \hline
  \end{tabular}
  \label{tab:swmsmonitoringfeatures}
\end{table*}

Table~\ref{tab:swmsmonitoringfeatures} shows the monitoring information the scientific workflow management systems (SWMS) offer themselves out-of-the-box.
In addition, various external tools and sources can be used independently.
\begin{itemize}
  \item Pegasus~\cite{pegasus} exclusively uses HTCondor as the resource manager, which in turn can use different infrastructure providers and architectures (such as Amazon EC2 or Open Science Grid). Monitoring data is submitted live into a relational database~\cite{deelmanEvolutionPegasusWorkflow2019} and can be queried and visualized from there (e.g., with ElasticSearch and Grafana).
  \item Nextflow~\cite{ditommasoNextflowEnablesReproducible2017} originated from the bioinformatics sector where common use cases are DNA analysis workflows~\cite{yates2021reproducible}. It gained more traction recently in other fields like processing and analysing satellite images~\cite{lehmann2021force} or magnetometer calibration~\cite{baderStypRekowski2022ICDMW}. Workflows in Nextflow use a domain-specific language based on Groovy. After a workflow execution finishes, Nextflow can compile reports containing the monitoring data.
  \item Airflow \cite{airflow} is tightly integrated into Python to describe workflows and the tasks they are made of. It can be configured to export some metrics by sending them to StatsD\footnote{https://github.com/statsd/statsd}, which can be used to further export them to Prometheus.
  \item Snakemake~\cite{snakemake} also uses a Python-based language to describe workflows and their tasks. Like in Nextflow, Snakemake can generate a report after the execution of a workflow. Furthermore, Snakemake supports 'panoptes' which is a server where monitoring data is sent to in real time by the SWMS. It provides an API to retrieve live data.
  \item Argo \cite{argo} is a workflow manager exclusively for Kubernetes. Workflows are described in YAML. It offers a web-based UI in which saved workflows can be viewed. The user can also access reports about previous workflow executions, both summarized as well as reports and logs of the executed tasks.
\end{itemize}

Most SWMS can be used with several resource managers.
However, most resource managers do not accept whole workflows to be submitted.
Therefore, the SWMS remains responsible for resolving the dependencies of the tasks and submitting them one after the other when their respective dependencies are satisfied.
Notable exceptions to this are the resource managers HTCondor and Slurm.
Slurm includes some mechanisms for job dependencies, while HTCondor uses the meta-scheduler DAGMan.

\subsection{Remarks and Discussion}

Table~\ref{tab:swmsmonitoringfeatures} shows that all SWMS offer some kind of workflow and task monitoring.
Especially monitoring on the workflow layer is widely supported.
The task monitoring layer yields bigger differences between the five SWMS.
However, apart from very fine-granular metrics, e.g., resource consumption for code parts, low-level task metrics, or fault diagnosis, the systems support most features.
A general observation was that the SWMS do not offer more detailed time-series data for a fine-grained analysis of the tasks' behavior.
The data is only available after a task's execution and cannot be retrieved during runtime.

A frequent technique to gather metrics used by the systems is to wrap the tasks that need to be executed into a bash script.
At the end of the execution, the script gathers metrics like the CPU time and memory usage from the dying process and writes them into a trace file that the SWMS then reads. 

Many state-of-the-art resource managers can not handle a complete workflow execution and instead rely on one-by-one task submissions from a separate SWMS.
Therefore, SWMS support one or even multiple resource managers in order to submit their tasks to them.
However, these resource managers provide no uniform programming interface.
Therefore, support for these monitoring layer would be very inconsistent depending on the chosen resource manager.
Our table reflects this observation as, except for the infrastructure status in Argo,  all systems do not provide metrics on this layer.

The monitoring from the machine layer is also not supported by any of the SWMS.
Since the resource managers are traditionally responsible for the actual task execution, they observe the cluster machines and their metrics.
Therefore, the gathering of machine monitoring metrics is handled by the resource manager.

Our table shows that the focus of monitoring information provided by SWMS clearly lies on the workflow layer. That is expected as their main responsibility is resolving workflow-specific dependencies, handling the task and data management of the workflow, and passing the execution to the resource manager.

Since an SWMS alone does not provide sufficient monitoring metrics, we argue for a uniform SWMS API interface that exchanges data with the resource manager.
First, this would help to provide additional data to the resource manager, e.g., the workflow DAG, to improve scheduling decisions~\cite{bader2021tarema,baderReshi2022IPCCC} or estimating the task runtimes~\cite{baderLotaruLocallyEstimating2022}.
Second, the user could directly access cluster and machine metrics collected by the resource manager on his system to help in debugging a workflow~\cite{tschuter2019top} or finding resource bottlenecks~\cite{bottlemod, will2022get, will2022ruya}.

%% file: sections/5_conclusion.tex
This paper presented an architectural blueprint for advanced monitoring of scientific workflows.
We classified monitored metrics into four monitoring layers: the resource manager, the workflow, the machine, and the task layer.  
These layers each define the monitoring subject of the corresponding monitoring metric.
The hierarchical order of the layers enables higher layers to access lower-layer metrics to operate.

We investigated five state-of-the-art scientific workflow management systems (SWMS) to classify their monitoring capabilities according to our four-layer-based monitoring approach.
Our analysis showed that while all the investigated SWMS provide some form of execution monitoring, especially on the workflow and the task layer, none supports monitoring at the machine layer.

Though our presented hierarchy is an accurate model, in practice, the workflow manager often submits task instances one-by-one to the resource manager, separating the workflow system and the resource manager.
Thus, the resource manager cannot access workflow-specific monitoring metrics, limiting the interaction and exchange of monitoring metrics and resulting in bad scheduling decisions. 
Better interaction between workflow and resource managers will be a key topic for future research.
Further, we want to use our proposed work to identify weaknesses in workflow monitoring systems and provide recommendations for more sophisticated workflow monitoring.